\documentclass[iicol]{sn-jnl}

\jyear{2021}%

\theoremstyle{thmstyleone}%
\theoremstyle{thmstyletwo}%

\theoremstyle{thmstylethree}%

\usepackage{color, soul}
\def\etal{\emph{~et~al. }}

\begin{document}

\title[Article Title]{Scalable Cellular V2X Solutions:	
Large-Scale Deployment Challenges of Connected Vehicle Safety Networks}


\author*[1]{\fnm{Ghayoor} \sur{Shah}}\email{gshah8@knights.ucf.edu}

\author[1]{\fnm{Mahdi} \sur{Zaman}}\email{mahdizaman@knights.ucf.edu}

\author[1]{\fnm{Md} \sur{Saifuddin}}\email{md.saif@knights.ucf.edu}

\author[1]{\fnm{Behrad} \sur{Toghi}}\email{toghi@knights.ucf.edu}

\author[1]{\fnm{Yaser} \sur{Fallah}}\email{yaser.fallah@ucf.edu}

\affil[1]{\orgdiv{Connected and Autonomous Vehicle Research Lab}, \orgname{University of Central Florida}, \orgaddress{\city{Orlando}, \state{Florida}, \country{USA}}}




\abstract{Vehicle-to-Everything (V2X) communication is expected to accomplish a long-standing goal of the Connected and Autonomous Vehicle (CAV) community to bring connected vehicles to roads on a large scale. A major challenge, and perhaps the biggest hurdle on the path towards this goal is the scalability issues associated with it, especially when vehicular safety is concerned. As a major stakeholder, 3rd Generation Partnership Project (3GPP) based Cellular V2X (C-V2X) community has long been trying to research on whether
vehicular networks are able to support the safety-critical applications in high-density vehicular scenarios. This paper attempts to answer this by first presenting an overview on the scalability challenges faced by 3GPP Release 14 Long Term Evolution C-V2X (LTE-V2X) using the PC5 sidelink interface for low and heavy-density traffic scenarios. Next, it demonstrates a series of solutions that address network congestion, packet losses and other scalability issues associated with LTE-V2X to enable this communication technology for commercial deployment. In addition, a brief survey is provided into 3GPP Release 16 5G New Radio V2X (NR-V2X) that utilizes the NR sidelink interface and works as an evolution of C-V2X towards better performance for V2X communications including new enhanced V2X (eV2X) scenarios that possess ultra-low-latency and high-reliability requirements.}




\maketitle


\section{Introduction}\label{sec1}
The adoption of Vehicle-to-Everything (V2X) communication is expected to accomplish a long-standing target of Connected and Autonomous Vehicle (CAV) community to assist in the safety and efficiency of Intelligent Transportation Systems (ITS){~\cite{cosgun2017towards}}. Significant time and effort have been spent on designing robust and reliable V2X communication technologies and a wide variety of Cooperative Vehicle Safety (CVS) applications relying on Vehicular Ad-Hoc Networks (VANETs) have been conceptualized \cite{ahmed2011vehicle}. However, the main challenge towards the potential commercial deployment of V2X communication has been the scalability issues associated with it. Cellular Vehicle-to-everything (C-V2X) \cite{molina2017lte} developed by 3rd Generation Partnership Project (3GPP) is the leading V2X communication technology in the United States due to its demonstrated superior performance in terms of latency, reliability, and communication range \cite{useof5.850}. Recently, the regulators have also been showing a keen interest in the prompt large-scale commercial deployment of C-V2X.

As a major stakeholder, the 3GPP based C-V2X community has long been 
researching whether C-V2X networks are capable of supporting the safety-critical applications in heavy-density traffic scenarios\mbox{~\cite{molina2017lte, bazzi2020wireless, toghi2019analysis, saifuddin2020performance, shah2020rve}}.
However, as analyzed in the afore-mentioned studies along with others\mbox{~\cite{toghi2019analysis, yoon2020balancing}}, and as experimentally presented later in this paper, C-V2X is prone to deteriorating performance under congested scenarios when employed using standard configurations as per 3GPP Release 14 Long Term Evolution C-V2X (LTE-V2X).
The focus throughout the paper is on LTE-V2X Mode-4 that uses the PC5 sidelink interface.
To resolve the scalability challenges with standard LTE-V2X Mode-4,
this paper provides novel solutions that can strongly enhance the performance of LTE-V2X in different traffic scenarios, e.g., extended frequency bandwidth, congestion control based on transmission rate control, and the one-shot based transmissions 
to avoid consecutive packet collisions as a consequence of bad resource allocation.
Utilizing these solutions is shown to significantly improve the scability performance of LTE-V2X in heavy-density traffic, thereby providing a path towards its mass commercial deployment.

Given the inherent limitations of LTE-V2X and the future research trends in the C-V2X community, this paper also provides a summary of 3GPP Release 16 5G New Radio C-V2X (NR-V2X) that can be used as an evolution of C-V2X. It shares the key features of NR-V2X that not only complement LTE-V2X in basic safety V2X cases but also assist in enhanced V2X (eV2X) cases that contain stringent latency and reliability requirements. Additionally, the current research status in the scalability aspect of V2X communication under NR-V2X is discussed in the paper.
Since the scalability-related research in NR-V2X is still under discussion amongst the C-V2X community, the paper does not provide an experimental comparison between LTE-V2X and NR-V2X and rather keeps its focus on the scalability aspects of LTE-V2X.

The rest of the paper is organized as follows. Sections \ref{sec2} and \ref{sec3} provide a high-level overview and performance details of LTE-V2X, respectively. In section \ref{sec4}, the paper presents the standardization efforts by the CAV community for LTE-V2X. Sections \ref{sec5} and \ref{sec6} provide the scalability solutions and closing remarks for LTE-V2X, respectively.
Section \ref{sec7} provides description and current research status of scalability with regards to NR-V2X. Finally, the paper is concluded in section \ref{sec8}.

\begin{figure}[b]
\centerline{\includegraphics[trim=0 0 0 0,clip,width=.48\textwidth]{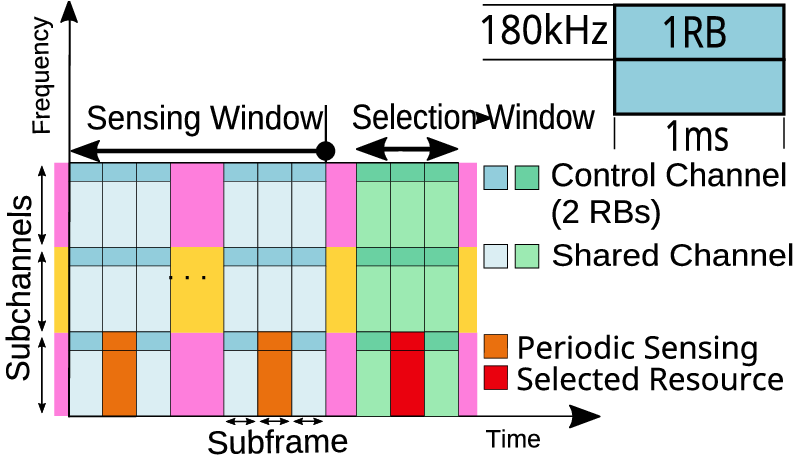}}
\caption{LTE-V2X Resource Allocation using SB-SPS}
\label{fig:SPS}
\end{figure}


\section{LTE-V2X Overview}\label{sec2}
3GPP Release 14 standard introduced V2X capability using Uu and PC5 sidelink \cite{molina2017lte}. We refer to this as LTE-V2X communication in contrast to 5G NR-V2X by Release 16. LTE-V2X enables Vehicular User Equipments (VUEs) to communicate in presence (Mode-3) and absence (Mode-4) of cellular network coverage and can be utilized to share situational awareness and information that enables safety applications; resulting in safer transportation and improved traffic efficiency, examples of which are advanced driver assistance systems (ADAS) such as lane-keeping assist (LKA) and cooperative adaptive cruise control (CACC), etc~\cite{vukadinovic20183gpp, shah2022enabling,shah2023arow}.

\subsection{MAC Layer}
LTE-V2X Media Access Control (MAC) and Physical (PHY) layers as well as their performance are studied thoroughly in prior arts~\cite{molina2017lte, toghi2018multiple}. Mode-4 adopts a distributed scheduling protocol to autonomously choose radio resources for transmission which is known as Sensing-based Semi-persistent Scheduling (SB-SPS). 

In the vanilla SB-SPS, each VUE keeps its own channel sensing history, using that as shown by the sensing window in Figure \mbox{\ref{fig:SPS}}, a VUE aims to allocate a radio resource that is less likely to be used by others. When a BSM transmission request is received from the upper layers at time $t$, the VUE monitors the sensing window ($[t-1000]ms$) and interpolates potential transmissions in the selection window ($[t+100]ms$). The periodic nature of BSM transmissions is helpful in interpolating and identifying the resources in the selection window that have a high probability of occupancy by any of the VUEs from whom a packet is successfully received in the sensing window. Any resource block with a Reference Signal Received Power (RSRP) over a certain threshold is considered as occupied. This identification allows the transmitting VUE to rank and sort the resource blocks in selection window in terms of the received power at those block locations. When at least 20\% of the total resources in the selection window are available for transmission, one of the available resources is reserved at random.

A BSM may require one or more resources for successful transmission. Hence for each transmission request, the SB-SPS procedure returns sufficient number of adjacent or non-adjacent resources within a single subframe. Several consequent resources in regular interval may also be reserved for a number of future transmissions. However, whether the future transmission opportunities will be reused or not is a function of a positive-biased probabilistic outcome. We redirect interested readers to\mbox{~\cite{toghi2018multiple, molina2017lte}} for a detailed description and analysis of the SB-SPS procedure.

The finite sensing window of each VUE is formed on basis of their perceived channel status. Hence under Mode-4 operation, the resource shortlist returned by SB-SPS is biased to the partial channel information at any given time. This perception can be corrupted by interference and noise, specially at congested traffic scenario where the number of resource candidates are higher than the number of finite available resources (as per the pigeonhole principle). In addition, the SB-SPS procedure do not allow re-evaluation of a resource after it's reservation, which can also increase the chance of interference at heavy-density traffic.

\subsection{PHY Layer}
LTE-V2X physical layer resource pool can be visualized as a two-dimensional resource grid along time and frequency as shown in Figure \ref{fig:SPS}. The time scale granularity is 1ms and is called a subframe. Each subframe is conceptually divided over several subchannels, where each subchannel consists of several resource blocks (RBs). A packet transmission can consume a specified collection of RBs over the same subframe, depending on the Modulation and Coding Scheme (MCS). A transmission is carried over two fundamental components: Signal Control Information (SCI), containing relevant radio control information over the control channel and Transport Block (TB) bears the actual packet data over the shared channel.

\begin{table}[t]
\setlength{\tabcolsep}{3pt}
\begin{center}
\caption{NS-3 Simulation Parameters}
\label{table:ns-3 sim params}
\begin{tabular}{l r}
\hline
Parameter & Value\\
\midrule
Packet Size & 190bytes\\
Tx Power (Baseline) & 23dBm\\
Tx Frequency (Baseline) & 10Hz\\
MCS index & 5\\
Carrier Frequency & 5.89GHz \\
CBR Threshold & -92dBm \\
Bandwidth & 10, 20MHz\\
Tx/Rx Antenna Height & 1.6m \\
Traffic Density (VUEs/100m) & 13.2 (Low), 83.2 (Heavy)\\
 Road Length (Highway) & 4.8km (Linear)\\
 No of Lanes & 6 \\ 
 Traffic Speed & 108km/h \\
 Propagation Loss Model & UCF I-405 Model \cite{ehsan_2020_channelmodel} \\
Simulation Duration & 100s\\
CBR Measurement Interval & 100ms\\

\botrule
\end{tabular}
\end{center}
\end{table}

\section{LTE-V2X Baseline Performance in Congested Scenarios}\label{sec3}
One of the main advantages attributed to LTE-V2X as compared to its predecessor, Dedicated Short-Range Communication (DSRC) \cite{kenney2011dedicated} is its extended range and a higher link budget~\cite{molina2017lte, saej2735}. Although some studies \cite{toghi2018multiple} showed that LTE-V2X proved to be a reliable mode of communication for low-density traffic scenarios, concurrent studies showed the vulnerability of LTE-V2X in heavy-density traffic~\cite{toghi2019analysis, yoon2020balancing}. This study demonstrates this using a simulation environment based on a state-of-the-art network simulator (NS-3). The simulation parameters used for the experiments are shown in Table \ref{table:ns-3 sim params}. A total of two traffic densities (low, heavy) are used to evaluate the performance of LTE-V2X.

The two main metrics used to measure the system-level performance of LTE-V2X for this study are:
\begin{itemize}
    \item Packet Reception Ratio (PRR): PRR is computed from the receiving VUE’s perspective as the ratio of successfully decoded and total Basic Safety Messages (BSMs) sent to the receiving VUE.
    In simpler terms, PRR quantifies the proportion of successfully received BSMs compared to the total number of transmitted BSMs. This study defines reasonable communicating pairs as all VUE pairs within a 1000m range, and calculate the PRR ratio across this set of VUE pairs. In an ideal wireless channel, PRR would be 100\%. However, in real-world scenarios, PRR can be reduced due to practical factors such as interference and noise in the wireless channel. In the experimental scenario of a bi-directional highway as in this study, the main cause of packet loss can be attributed to interference from other sidelink transmissions.
    \item Information Age (IA): IA is computed as the average time-gap between two consecutive BSMs from the same transmitter. IA performance is presented for two distance bins (0-200m, 200-300m) to measure the behavior in lower and relatively higher VUE-VUE ranges, respectively. In this paper, IA is computed as a complimentary CDF (CCDF) plot in the logarithmic scale.
    In C-V2X systems, achieving low IA values is crucial for enabling timely and accurate decision-making processes, such as collision avoidance and traffic management. The objective of minimizing IA is to ensure that VUEs have access to the most recent (and hence relevant) information and enhance the overall effectiveness and safety throughout the network.
\end{itemize}

\begin{figure}[t]
\centerline{\includegraphics[trim=20 20 20 10,clip,width=.48\textwidth]{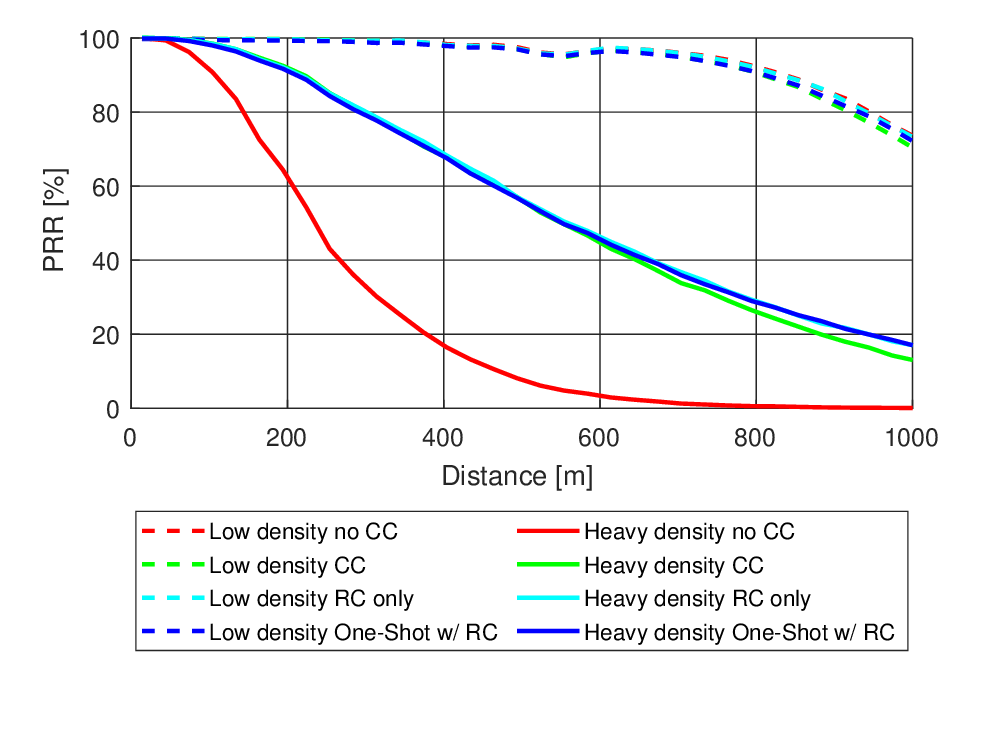}}
\caption{PRR for different congestion control schemes over varying traffic density. no CC denotes no Congestion Control applied, CC denotes Congestion Control with Power and Rate control according to SAE J2945/1, RC only denotes Congestion Control with only Rate Control according to SAE J3161/1 without One-Shot, and One-Shot w/ RC denotes Congestion Control with Rate Control along with One-Shot according to SAE J3161/1.}
\label{fig:PRR-combined}
\end{figure}

\begin{figure}[t]
\centerline{\includegraphics[trim=30 20 50 30,clip,width=0.48\textwidth]{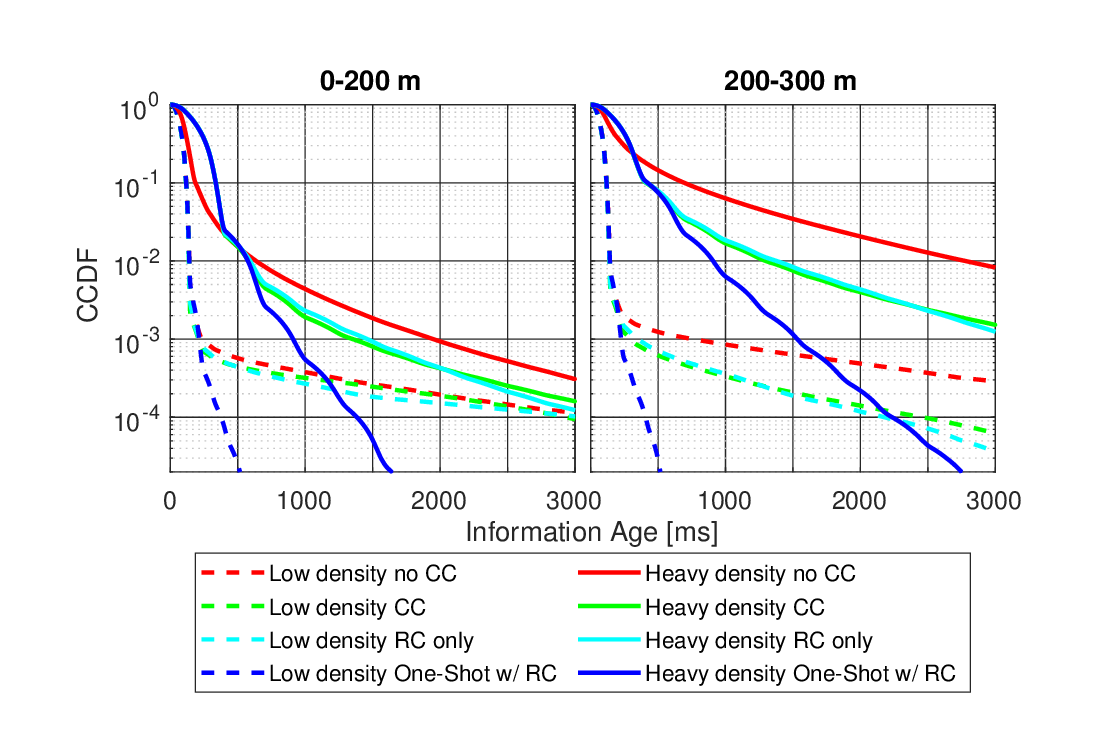}}
\caption{IA for different congestion control schemes over varying traffic density and distance range. Lower Value of IA denotes a better overall performance.}
\label{fig:IA-combined}
\end{figure}

Figures \ref{fig:PRR-combined} and \ref{fig:IA-combined} show the PRR and IA, respectively, to compare the performances of baseline LTE-V2X without Congestion Control (no CC) 
with proposed scalability solutions such as
LTE-V2X with Congestion Control comprising of Power and Rate Control (CC), LTE-V2X with Congestion Control containing Rate Control only (RC only), and LTE-V2X with Congestion Control including Rate Control only with One-Shot based transmissions (One-Shot w/ RC). 
Focus in this section is on baseline LTE-V2X without Congestion Control (no CC). It should be noted that baseline LTE-V2X refers to LTE-V2X Mode-4 using standard configurations as per 3GPP Release 14. The proposed solutions and their plots in Figures \mbox{\ref{fig:PRR-combined}} and \mbox{\ref{fig:IA-combined}} are explained in later sections.

As illustrated in the PRR plots in Figure \mbox{\ref{fig:PRR-combined}}, when VUE pairs are in close proximity within 100m of each other, the PRR is consistently above 90\% for both low and heavy-density traffic scenarios. However, it significantly decreases as the distance between the VUEs increases, specially in heavy-density traffic. In the case of low-density traffic, PRR remains relatively high even for VUE pairs as far apart as 800m. Contrarily, in heavy-density scenarios, PRR exhibits a continuous downward trend. Even VUE pairs with distance of 200m can experience a PRR as low as 60\%, as illustrated in Figure \mbox{\ref{fig:PRR-combined}}.


On the other hand, the impact of IA, as shown in Figure \mbox{\ref{fig:IA-combined}} is different for VUE pairs at different distances. The closer two VUEs are, the more crucial it is to maintain low latency (and thus low IA) within their communication. Hence we observe IA separately for two groups of VUE pairs, (1) close proximity (0-200m separation, left subplot of Figure \mbox{\ref{fig:IA-combined}}) and (2) moderate proximity (200-300m, right subplot of Figure \mbox{\ref{fig:IA-combined}}). Since the impact of IA carries more significance for (1), a low IA in this group is more devastating than in (2).
Observing Figure \mbox{\ref{fig:IA-combined}} for low-density traffic, the performance is similar to the proposed congestion control plots at (0-200m) range, whereas it deteriorates in (200-300m) range. For heavy-density, it can be observed that IA is higher than other congestion control schemes for both (0-200m) and (200-300m) VUE-VUE ranges. Given the congestion issues faced by baseline LTE-V2X as explained above, it became inevitable for relevant groups to convene on a potential standardization of an LTE-V2X congestion control algorithm before commercial deployment.

\section{Standardization Efforts for C-V2X Congestion Control}\label{sec4}
The Society of Automotive Engineers (SAE) provides system standards for V2X communication in USA.  Previously, SAE standardized VANET protocols based on IEEE 802.11p specifications, that includes: (1) SAE J2735: defines the content of the periodic BSMs to be shared through the V2X RAN \cite{saej2735}, (2) SAE J2945/1: provides a set of instructions and procedures for DSRC MAC, including congestion control management \cite{saej2945}.

C-V2X supports the J2735 standard for its own periodic messages. However, for the latter, study on retrofitting J2945/1 congestion control procedures to LTE-V2X \cite{toghi2018multiple} turned out to be unsatisfactory and incomparable due to the inherent lower layer differences between the two technologies. Although LTE-V2X failed to perform optimally with the imported settings, the studies showed potential that further tuning and optimization on the J2945/1 algorithm may enable efficient congestion management in C-V2X. In this regard, SAE J3161/1 \cite{saej3161} aimed to develop protocols with similar goal as J2945/1 to aptly fit with C-V2X.

To maintain Quality of Service (QoS) at heavy-density traffic, research efforts are undergoing through various approaches. This includes amendment suggestions on MAC as well as potential benefits from a larger physical spectrum. The following section shares some of the most impactful insights on the effect of these modifications as discovered by our workgroup and other researchers.

\section{Scalability Solutions for LTE-V2X Congested Scenarios}\label{sec5}
Given the challenges associated with baseline LTE-V2X as shown in Section \mbox{\ref{sec3}}, this section proposes novel scalability solutions to improve the performance of LTE-V2X in congested scenarios. Main focus is on frequency bandwidth extension, congestion control through transmission rate and power, and reducing consecutive packet collisions through one-shot based transmissions.

\begin{figure}[t]
\centerline{\includegraphics[trim=0 0 10 30,clip,width=0.48\textwidth]{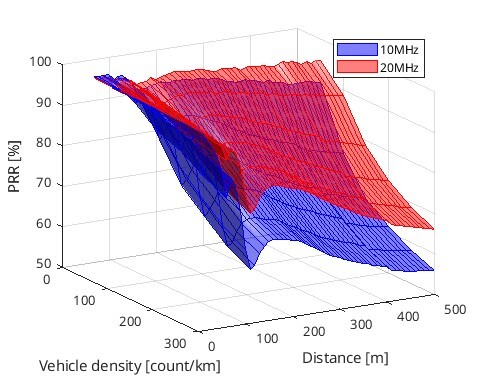}}
\caption{PRR for 10MHz and 20MHz bandwidth in C-V2X}
\label{fig:PRR_10_20MHz}
\end{figure}

\subsection{Frequency Bandwidth Extension}\label{subsec5.1}
In 1999, Federal Communications Commission (FCC) reserved 75MHz of spectrum in 5.9GHz band for ITS. Unlike DSRC, C-V2X is not limited in frequency bandwidth since its segmented resource structure has minimal impact from Doppler effect. So, it can reduce channel congestion by adding more bandwidth for its basic safety applications. To the benefit of C-V2X, FCC proposed in November 2020 to reserve 30MHz of the 5.9GHz band for C-V2X \cite{useof5.850}. 
Using parameters in Table \mbox{\ref{table:ns-3 sim params}}, this study compares the PRR using 10MHz and 20MHz bandwidth under variable VUE-VUE distances and traffic densities as shown in Figure \mbox{\ref{fig:PRR_10_20MHz}}. It can be clearly observed for all traffic densities and distances that a bandwidth of 20MHz significantly outperforms 10MHz in terms of PRR.

\subsection{Congestion Control}\label{subsec5.2}
As explained earlier, one of the primary hurdles ahead of the mass-adoption of V2X communication technologies is the performance challenges under heavy-density traffic. Communication technologies usually employ two main approaches to determine network congestion, namely channel utilization-based sensing \cite{huang2010adaptive} and message relevance-based assessment \cite{kosch2006scalability}. In general, a VUE obtains situational awareness via both periodic BSMs as well as aperiodic event-triggered messages with different channel access priorities and structure \cite{toghi2018multiple}. For BSMs, the Decentralized Congestion Control (DCC) algorithm for DSRC under SAE J2945/1 standard \cite{saej2945} aims to reduce collision probability by utilizing a combination of these two approaches into an algorithm using two levers: (1) adapting transmission power to maintain Channel Busy Ratio (CBR), and (2) adapting transmission rate to reduce channel load without compromising latency performance, as summarized in Figure \ref{fig:Control Characteristics Power, Rate}.

\begin{figure}[b]
\centerline{\includegraphics[trim=0 0 10 10,clip,width=0.48\textwidth]{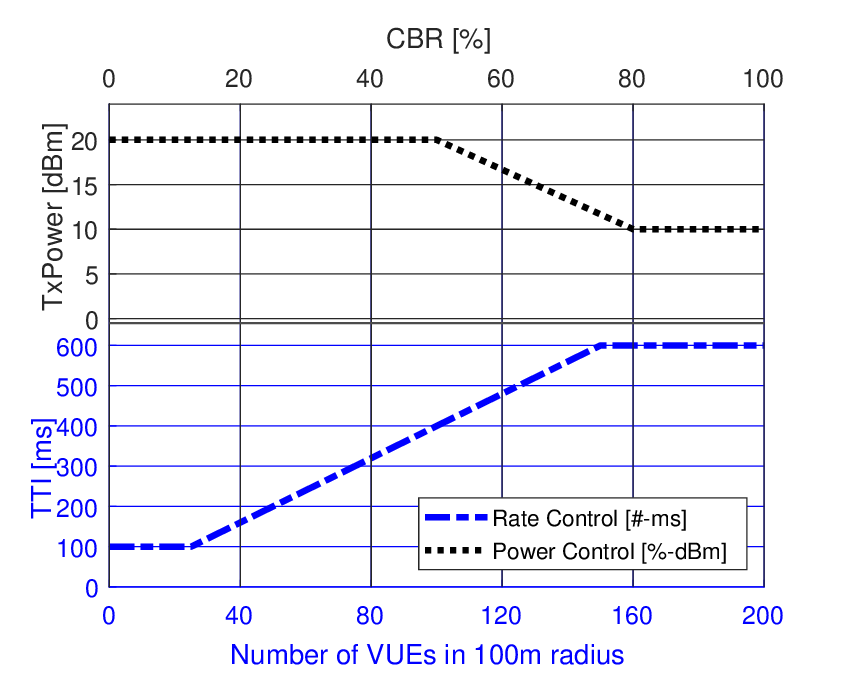}}
\caption{Summary of control characteristics for Power and Rate Congestion Control. TTI denotes Transmission Time Interval}
\label{fig:Control Characteristics Power, Rate}
\end{figure}

LTE-V2X configuration identical to DCC for DSRC yields longer range and enhanced reliability in its default setup, but the increased interference at heavy-density traffic emphasizes on the need for a DCC scheme configured to fit LTE-V2X operations. In this effort,~\cite{toghi2019analysis,yoon2020balancing} demonstrate the dominating effect of transmission rate control over power control. 
Figure \mbox{\ref{fig:Control Characteristics Power, Rate}} also demonstrates this by providing a comparative activation of rate control and power control as designed in J2945/1 standard\mbox{~\cite{saej2945}}. Along x-axis are the control signals in terms of traffic density (bottom) and CBR (top). At low-density traffic ($\leq$ 30veh/100m), congestion control procedures remain dormant. At around 30veh/100m, the rate control procedure is activated to relax the periodic transmission interval to adjust for increased traffic density. VUEs perceive the notion of traffic density by the reception record. Once the density is perceived to be over 30veh/100m, each VUE employs their transmission rate adjustments accordingly. On the other hand, power control activation is a function of the CBR, which remains unchanged at this density due to early activation of rate control as shown in Figure \mbox{\ref{fig:Control Characteristics Power, Rate}}.

As traffic density increases, rate control continues to relax the transmission interval and keeps CBR unaffected until the density reaches 100veh/100m. At this point, power control is also employed to address further congestion.
However, power control activation introduces a trade-off regarding the transmission range. The successful reception of a packet relies on the Signal-to-Interference-plus-Noise Ratio (SINR), which is determined by the transmission power and varies according to the channel and scenario in consideration. Consequently, reducing the transmission power results in a compromise in the maximum reception range. Thus, in a vehicular wireless medium, it is advantageous to explore alternative methods to address congestion while maintaining a long transmission range. While power control adjustment was previously critical for congestion control in the DSRC channel, Figure \mbox{\ref{fig:Control Characteristics Power, Rate}} indicates that for C-V2X, adopting a more sophisticated rate control approach while disregarding power control could potentially enhance the overall system performance.

This study also shows this phenomenon by running simulations utilizing three congestion control schemes: Congestion Control comprising of power and rate Control (CC), with rate control only (RC only), and with only rate control with one-shot based transmissions (One-Shot w/ RC). This subsection focuses on the first two schemes, one-shot discussion and plots are discussed in the next subsection. On a high-level, it can be observed in both Figures \mbox{\ref{fig:PRR-combined}} and \mbox{\ref{fig:IA-combined}} that plots of congestion control with rate and power control (CC) and congestion control with rate control only (RC only) are similar for both traffic densities. This confirms the earlier-made observation from Figure \mbox{\ref{fig:Control Characteristics Power, Rate}} that rate control is sufficient for congestion control in LTE-V2X.

Congestion control efforts target resource-level correction by randomizing and spreading the resource allocation over the resource pool, and thereby reducing collision probability at each resource. At low-density traffic, when the channel bandwidth offers higher resource availability than the demand by the active VUE count, each VUE can flexibly select a resource from the available resource pool for their respective transmissions. Hence at low-density traffic, congestion control mechanisms are not expected to yield distinctive advantage. Our simulation results in Figures \ref{fig:PRR-combined} and \ref{fig:IA-combined} affirm this notion.

However, congestion control mechanisms thrive at maintaining network quality at heavy-density traffic where resources are scarce. In Figures \ref{fig:PRR-combined} and \ref{fig:IA-combined}, we observe noticeable improvement in PRR and IA with congestion control schemes (CC, RC only) as compared to baseline LTE-V2X (no CC). In terms of PRR, VUE pairs within (0-200m) range enjoy up to 50\% increase in reception using congestion control schemes as compared no congestion control. Improved PRR also increases the frequency of message reception, thereby reducing latency. IA plots in Figure \ref{fig:IA-combined} capture this reduction in latency. We observe that the congestion control schemes (CC, RC only) reduce IA for both (0-200m) and (200-300m) VUE-VUE ranges. For (0-200m) range, 99.9\% ($CCDF = 10^{-3}$) VUEs enjoy at least 500ms lower IA compared to when no congestion control (no CC) is applied. The improvement is more significant for (200-300m) range, where 99\% ($CCDF = 10^{-2}$) VUEs enjoy at least 1300ms lower IA than no congestion control (no CC) under any presented congestion control setting.

\subsection{One-Shot Based Transmissions}\label{subsec5.3}
Since LTE-V2X uses SB-SPS to allocate resources for periodic transmissions, there can be prolonged spells of persistent packet losses among colliding nodes due to bad resource allocation\mbox{~\cite{SAIFUDDIN2023}}. This phenomenon is more prevalent in heavy-density scenarios. 
Bazzi\mbox{~\etal~\cite{bazzi2020wireless}} analyzed this problem as `wireless blind spot' scenario and proposed a time-limit based resource usage to reduce the duration of bad resource allocation. While the adoption of this technique reduces blind spot problem, this strict limitation on `maximum keep time' can affect the overall reception probability as the performance of reception increases when the `maximum keep time' is higher, as shown in\mbox{~\cite{toghi2018multiple}}.

In this regard, Fouda\mbox{~\etal~\cite{fouda2021interleaved}} proposed an alternative proposition known as the one-shot based transmission. This proposition limits consecutive periodic transmissions at an allocated periodic resource, thereby reducing the number of consecutive packet losses. To achieve this, it introduces a separate singular aperiodic resource allocation for packet transmission. This allocation happens  probabilistically every  2 to 6 transmissions, which is decided using a counter called `one-shot counter'. As this counter expires, the transmitting VUE avoids the possible colliding resource by using this one-off resource allocated using SB-SPS. After this transmission, the regular periodic transmission resumes on the originally allocated resource from the next packet. This one-shot based transmission algorithm works alongside the rate control  algorithm and successfully shortens the tail of IA distribution. This method thus avoids disastrous network failure due to complete communication outage between specific VUEs. This work was eventually adopted by SAE in J3161/1 application\mbox{~\cite{saej3161}}. The same method has been adopted in the one-shot simulations in this study.

Figures \ref{fig:PRR-combined} and \ref{fig:IA-combined} represent the PRR and IA performance of one-shot based transmissions (One-Shot w/ RC). It can be noted that the PRR with this setting is very similar to that of the other congestion control schemes (CC, RC only), which shows that the reception is not affected by the additional one-off transmission using one-shot. In terms of IA, one-shot based transmissions (One-Shot w/ RC) outperform the other congestion control schemes. The most significant improvement is observable in the (200-300m) range in Figure \ref{fig:IA-combined}, where IA is reduced by more than 1500ms for 99.9\% ($CCDF = 10^{-3}$) VUEs as compared to other congestion control schemes (CC, RC only). For heavy-density traffic, the IA plots for one-shot can be seen to perform slightly worse than other schemes at lower percentile ($CCDF = 10^{-1}$), since one-shot adds to the channel uncertainty. In contrast, it significantly reduces the rare, prolonged collisions as in the lower tail-end IA CCDF values as opposed to other congestion control schemes.


\begin{table*}[t]
\caption{A high-level difference in the features of LTE-V2X and NR-V2X}\label{table:LTE vs NR features}%
\resizebox{\textwidth}{!}{\begin{tabular}{@{}lll@{}}
\toprule
\textbf{Feature} & \textbf{LTE-V2X (Release 14)}  & \textbf{NR-V2X (Release 16)}\\
\midrule
Waveform & SC-FDMA & CP-OFDM\\
Numerology & 1ms subframe & 0.125, 0.25, 0.5, or 1ms slot\\
Sub-Carrier Spacing & 15KHz & FR1: 15, 30, 60KHz, FR2: 60, 120KHz\\
Transmission Type & Broadcast & Broadcast, Groupcast, Unicast\\
DMRS & 4/subframe & Flexible\\
MCS & QPSK, 16-QAM & QPSK, 16-, 64-, 256-QAM\\
Feedback Channel & None & PSFCH\\
HARQ re-transmission & Blind & Blind or Feedback-based\\
SCI & Single-stage, separate RBs & Two-stage, shared RBs with TB\\
Maximum Channel Bandwidth & 20MHz (PC5), 40MHz (Uu-PC5) & 40MHz (PC5), 60MHz (Uu-PC5)\\
Distributed Resource Allocation & Mode-4 (SB-SPS) & Mode-2 (Dynamic, SB-SPS)\\
Resource Reservation Interval & 0, 20, 50, 100, n*100, $n = \{x:x\in \mathbb{N}, 1 \leq x \leq 10\}$ & 0, 1:99, 100, n*100, $n = \{x:x\in \mathbb{N}, 1 \leq x \leq 10\}$\\
\botrule
\end{tabular}}
\end{table*}

\section{Closing Argument for LTE-V2X}\label{sec6}
As explained in the previous sections, it can be observed that the performance of LTE-V2X can highly benefit from congestion control proposals for basic V2X use-cases. However, with eV2X use cases such as vehicle platooning, advanced driving, extended sensors, and remote driving \cite{3gppenhancement}, and the stringent latency requirements \cite{garcia2021tutorial} for such applications, LTE-V2X is not optimized for  meeting such requirements. For example.  these eV2X applications can require latency to be as low as 3ms, reliability to be as high as 99.999\%, transmission frequency to be as high as 100Hz, and packet sizes to be as large as 16KB. But, at a slot size of 0.5ms and requiring at least 2 slot for a packet transmission in regular cases, meeting this latency requirement while maintaining scalability is not supported. Moreover, unicast and groupcast transmissions benefit from efficient state information feedback which is not present in LTE-V2X.  This encouraged 3GPP to utilize a new air interface based on NR that contained several enhancements explained in the next section that could not only complement LTE-V2X in general, but also serve in eV2X use-cases. 



\section{3GPP Release 16: New Radio in C-V2X}\label{sec7}

\begin{figure}[t]
\centerline{\includegraphics[trim=20 0 80 10,clip,width=0.48\textwidth]{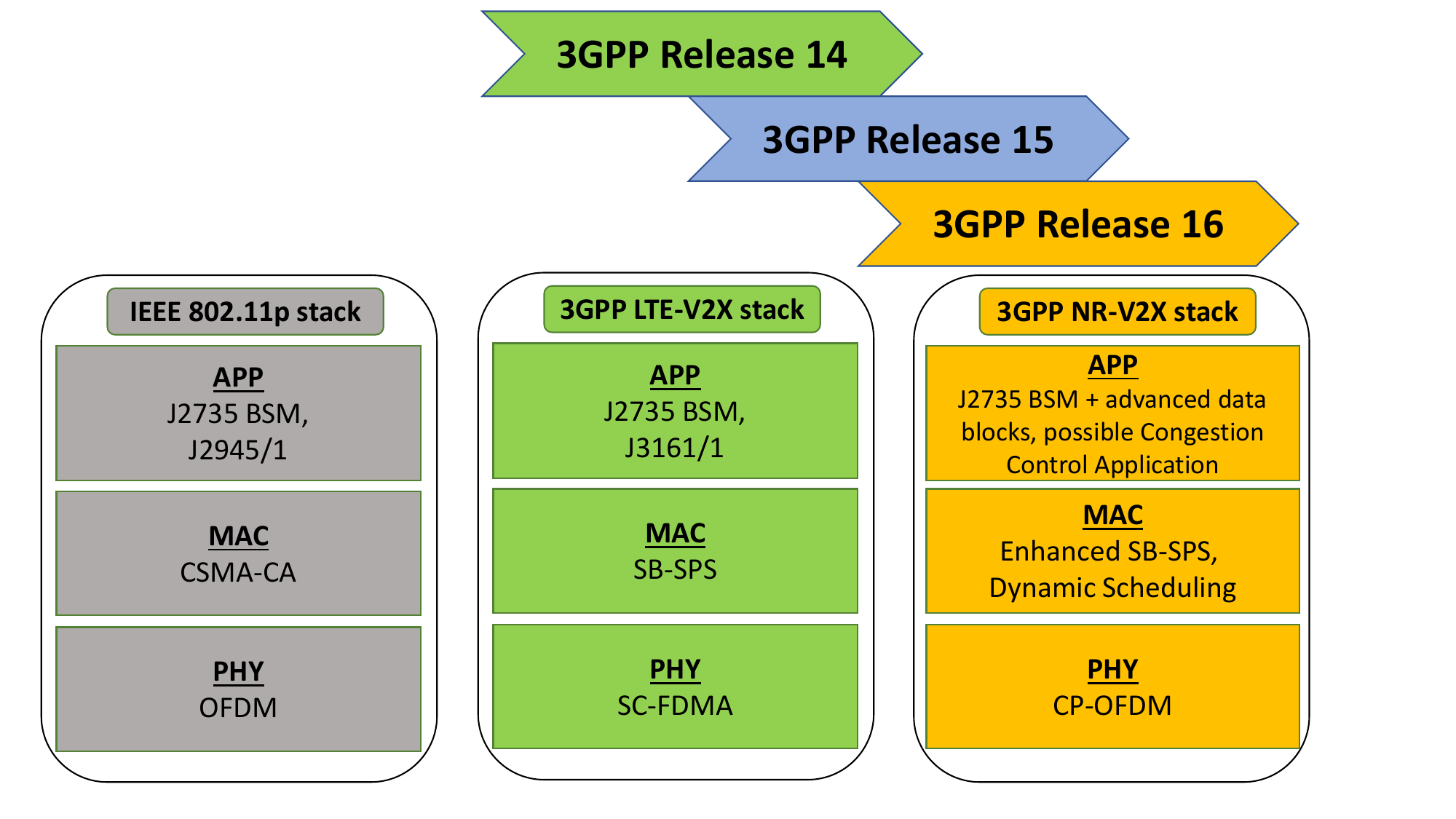}}
\caption{Progress of 3GPP C-V2X stack}
\label{fig:C-V2X evolution}
\end{figure}

There is a significant on-going effort amongst the 3GPP community to further improve reliability and scalability of LTE-V2X. In this regard, NR-V2X, based on 3GPP Release 16 \cite{NR:phy} is expected to address a plethora of issues including scalability concerns~\cite{garcia2021tutorial, bazzi2021design}. 
Figure \ref{fig:C-V2X evolution} provides a visual representation of the evolution of 3GPP from Release 12 to Release 16.
Table \ref{table:LTE vs NR features} provides a high-level difference between LTE-V2X and NR-V2X.
It can be observed from Table \mbox{\ref{table:LTE vs NR features}} that NR-V2X provides enhancements to most of the features of LTE-V2X and thus acts as an evolution of LTE-V2X that is capable of not only outperforming it in basic V2X use-cases, but also providing a platform for eV2X cases. Since NR-V2X is still a relatively new technology, there are several aspects related to scalability that are still under discussion amongst the research community. Therefore, this section aims to provide a summary of the components of NR-V2X sidelink communication standard introduced in Release 16 that can potentially improve scalability as compared to LTE-V2X. Although there is sparse literature on this topic, this section provides reference to preliminary studies that have aimed to analyze the impact of NR-V2X in scalability as compared to LTE-V2X.

\subsection{Physical Layer}\label{subsec7.1}

\subsubsection{Flexible Numerology and Bandwidth Part}\label{subsec7.1.1}

LTE-V2X is expected to solely operate on 5.9GHz band. But to scale for eV2X cases, NR-V2X allows the usage of two frequency ranges: (Frequency Range 1 (FR1) and Frequency Range 2 (FR2)). FR1 refers to the spectra between 410KHz and 7.125GHz, whereas FR2 refers to spectra between 24.25GHz and 52.6GHz. 
Although NR-V2X supports both frequency ranges, the design of NR-V2X is mainly based on FR1 and no specific optimization has been performed for FR2 yet, except for preliminary literature on addressing phase noise at such high frequency range\mbox{~\cite{zou2016impact}}. 
Thus, most recent studies working on NR-V2X are utilizing FR1\mbox{~\cite{saad2021advancements,ali2021impact,ali20213gpp}}.
Different carrier frequencies can cause different levels of multi-path fading which adds to the complexity of successful message reception. To mitigate this issue, NR-V2X allows the use of flexible numerology option, allowing different maximum carrier bandwidths, noticeably higher than in LTE-V2X. Major benefit of scalable numerology comes from a larger Sub-Carrier Spacing (SCS) that assumes a smaller slot, which translates to a lower latency. Secondly, a larger SCS provides a better protection against Doppler effects at high VUE speed. Also, a higher carrier bandwidth can be beneficial for large payloads generated by eV2X use-cases.

In addition, to support services that does not use large carrier bandwidths, NR-V2X incorporates sidelink Bandwidth Part (BWP) initially introduced in Release 15 NR Uu. Based on the chosen numerology, a VUE can be configured to use a SL BWP within the carrier bandwidth which is basically a continuous portion of bandwidth within the current carrier bandwidth with a singular numerology. 
Ali\mbox{~\etal~\cite{ali2021impact,ali20213gpp}} conduct preliminary analysis on flexible numerology and conclude that based on a fixed resource selection window, an increasing SCS contributes to a better Packet Inter-reception Delay (PIR) and lesser percentage of consecutive packet collisions that can arise as a consequence of bad resource allocation.
In another study, Campolo\mbox{~\etal~\cite{campolo20195g}} demonstrates the superiority of flexible numerology of NR-V2X over fixed numerology of LTE-V2X. The study provides results showing that NR-V2X outperforms LTE-V2X in PRR in all kinds of traffic density, BSM transmission frequency, and MCS.

\subsubsection{Feedback-based Retransmissions}\label{subsec7.1.2}
LTE-V2X sidelink radio, as defined in 3GPP Release 14, is solely based on broadcast communication. On the other hand, Release 16 allows unicast and groupcast communication for eV2X use-cases that require ad-hoc and targeted transmissions among VUEs. In Release 14, due to a sub-optimal resource allocation or a congested scenario, a transmitted packet may not reach the intended recipients. To counter this, Release 14 allows a single blind retransmission called Hybrid Automated Repeat Request (HARQ) after every periodic transmission without any acknowledgement signal from the receiver. Although this retransmission increases the chances of packet reception, it drastically increases the channel load, especially in congested scenarios. Therefore, to increase the reliability and scalability of C-V2X, Release 16 allows feedback-based retransmissions for unicast and groupcast communications. After every unicast or groupcast transmission, the receiver either sends an acknowledgement (HARQ-ACK) or negative acknowledgment (HARQ-NACK) upon successful/unsuccessful decoding of a TB. The transmitter then performs a retransmission of the original packet in the case of receiving a HARQ-NACK or if it does not receive any acknowledgement within a specified time. This HARQ feedback is sent from the receiving VUE via Physical Sidelink Feedback Channel (PSFCH). Thus, using these feedback-based retransmissions not only assist in reliability but also assist in scalability since it avoids unnecessary retransmission unless required \cite{ali20213gpp}.

\subsubsection{Two-Stage SCI}\label{subsec7.1.3}
NR-V2X introduces two-stage SCI to simplify SCI decoding and mitigate radio collisions. It splits the SCI information into two parts: stage-1 and stage-2 SCI. The smaller part, stage-1 SCI, includes resource allocation information, e.g., the address of radio resource containing the TB, message priority, and decoding information related to stage-2 SCI. Stage-2 SCI contains the remaining information such as MCS, DMRS locations, HARQ, transmitter/target VUE ID (for unicast and groupcast) etc. Stage-1 SCI is transmitted using Physical Sidelink Control Channel (PSCCH) whereas stage-2 SCI and TB is transmitted using Physical Sidelink Shared Channel (PSSCH).

\subsubsection{Physical Sidelink Broadcast Channel (PSBCH)}
PSBCH carries sidelink synchronization information
related to packet transmission. The information contained
within PSBCH includes a direct frame number which
allows a VUE to synchronize its radio transmissions
according to the sidelink timing reference. Secondly,
PSBCH also includes the 1-bit in-coverage indicator
which provides information about whether a
synchronization reference VUE is within network
coverage. Finally, PSBCH also contains the indications of
the slots and symbol-level time resources that can be
included in the resource pool.

\subsection{Resource Allocation}\label{subsec7.2}
In LTE-V2X Mode-4, SB-SPS enables VUEs to autonomously select their sidelink radio resources. On the other hand, NR-V2X in Release 16 Mode-2 broadens the scope by incorporating a dynamic scheduling approach along with a modified SB-SPS. The dynamic scheme allows resource selection for each transmission and subsequent re-transmission (HARQ) of a singular TB, if necessary. The inter-transmission time in the new SB-SPS is also flexible due more allowed values of Resource Reservation Interval (RRI) as shown in Table \ref{table:LTE vs NR features}. This allows for efficient usage of resources by VUEs based on their applications.

For both dynamic and SB-SPS schemes, another significant improvement in NR-V2X is in terms of the re-evaluation and pre-emption strategies~\cite{NR:phy, molina2022does}. Re-evaluation mechanism allows a VUE to re-evaluate the status of a previously selected resource prior to the scheduled transmission of its next packet. If the previously selected resource does not qualify to be in the shortlisted subset from the new window, it can be considered that this resource has been utilized by another VUE, and the VUE can reschedule its transmission from the newly gained shortlist of candidate resources and thus avoiding radio collision.

Pre-emption is a technique that assists in re-evaluation procedure by enabling a re-distribution of resources to facilitate high-priority traffic. During re-evaluation, a VUE keeps track of the packet priority for each reception. If re-evaluation suggests that the selected resource has also been recently occupied by another VUE, the VUE pre-empts on the basis of packet priority and vacates the selected resource if the other VUE’s packet has a higher priority. In this case, the VUE allocates a new resource from the new short-listed pool.

A few studies have been conducted comparing NR-V2X and LTE-V2X in FR1 based on the advancements in NR-V2X resource allocation as shown in this subsection.
Saad\mbox{~\etal~\cite{saad2021advancements}} compare NR-V2X in FR1 with SCS of 30 and 60 KHz against LTE-V2X and shows a promising improvement in Packet Delivery Ratio (PDR).
Similarly, a recent study by Shin\mbox{~\etal~\cite{shin2023vehicle}} provides a marked PRR improvement for NR-V2X when compared to LTE-V2X as a result of its resource allocation using re-evaluation and pre-emption. It also presents some insights into the usage of novel inter-VUE coordination (IUC) in conjunction with pre-emption and re-evaluation that can potentially resolve the hidden node problem.

\subsection{Future Trends: Application QoS and Congestion Control in NR-V2X}\label{subsec7.3}
NR-V2X defines similar access layer congestion control to its predecessors, based on Channel Occupancy Ratio (CR) and CBR which is reactive in nature~\cite{garcia2021tutorial, bazzi2021design}. However, unlike LTE-V2X, it does not directly address any scalability issues at higher network load. This field of research is still of ongoing interest from the CAV community. Recent study by McCarthy~\etal~\cite{mccarthy2022adapting} attempts to evolve earlier LTE-V2X DCC mechanism for NR-V2X from a MAC scheduling perspective. Although this allows an adaptive congestion control, no known published work on application QoS level assessment of the congestion control is available at the time of writing to the best of authors’ knowledge.

\section{Conclusion}\label{sec8}

This study focuses on the large-scale deployment challenges associated with LTE-V2X. To counter the performance deterioration of LTE-V2X in heavy-density traffic, the research community and automakers have suggested scalability issues for a more reliable usage of LTE-V2X in dense-traffic situations. The main propositions include an extended frequency bandwidth allowance, a fine-tuned congestion control algorithm for LTE-V2X medium with an emphasis on rate control, and finally the introduction of one-shot based transmissions to avoid persistent collisions. This paper runs simulations to prove the benefits of utilizing these proposals. Finally, the paper also discusses NR-V2X technology that can not only further improve upon the performance of LTE-V2X in basic V2X cases but can also handle enhanced V2X applications. Further research in NR-V2X can improve the safety and reliability of V2X communication.

\section*{Declarations}
On behalf of all authors, the corresponding author states that there is no conflict of interest.

\bibliography{sn-bibliography}
\bibliographystyle{unsrt}


\end{document}